\documentclass{llncs} 
\newtheorem{defn}{D\'efinition}
\newtheorem{prop}[defn]{Proposition}
\newtheorem{thm}[defn]{Theorem}
\newtheorem{lem}[defn]{Lemma}
\newtheorem{cor}[defn]{Corollary}
\newtheorem{con}{Conjecture}

\title{An efficient algorithm finds noticeable trends and examples
concerning the \v{C}erny conjecture}

\date{}
\author{A.N. Trahtman\thanks{Email: trakht@macs.biu.ac.il}\thanks{this version 
differs of LNCS version. Here an untrue lemma is omitted}}
\institute{Bar-Ilan University, Dep. of Math., 52900, Ramat Gan, Israel}
\begin{document}

\maketitle
\centerline{Lectures Notes in Computer Science, 4162(2006), 789-800}
\begin{abstract}
  A word $w$ is called synchronizing (recurrent, reset, directed)
word of a deterministic finite automaton (DFA)
if $w$ sends all states of the automaton on a unique state.
Jan \v{C}erny had found in 1964 a sequence of $n$-state complete
DFA with shortest synchronizing word of length $(n-1)^2$. He had
conjectured that it is an upper bound for the length of the
shortest synchronizing word for any $n$-state complete DFA.
\\
The examples of DFA with shortest synchronizing word of length
$(n-1)^2$ are relatively rare. To the \v{C}erny sequence were added
in all examples of \v{C}erny, Piricka and Rosenauerova (1971), of Kari (2001)
and of Roman (2004).
\\
  By help of a program based on some effective algorithms, a wide class of
 automata of size less than 11 was checked. The order of the algorithm finding
 synchronizing word is quadratic for overwhelming majority of known to date
 automata.
Some new examples of $n$-state DFA with minimal synchronizing word
of length $(n-1)^2$ were discovered. The program recognized some
remarkable trends concerning the length of the minimal
synchronizing word.
 \\
http://www.cs.biu.ac.il/$\sim$trakht/Testas.html.
\end{abstract}
 \section*{Introduction}
  We consider a DFA with complete state
 transition graph $\Gamma$ and transition semigroup $S$ over alphabet $\Sigma$.
Let $n$ be the size of DFA and $q$ be the size of $\Sigma$.
\\
The problem of synchronization of DFA is natural and various aspects
of this problem were touched upon the literature.
Synchronization makes the behavior of an automaton resistant
against input errors since, after detection of an error, a
synchronizing word can reset the automaton back to its original
state, as if no error had occurred. Therefore different problems of
 synchronization draw the attention.
\\
A problem with a long story is the estimation of the minimal length of
synchronizing word. Most known as a \v{C}erny conjecture, it was aroused
 independently by distinct authors.
  Jan \v{C}erny had found in 1964 \cite {Ce} $n$-state complete
DFA with shortest synchronizing word of length $(n-1)^2$ for $q=2$. He had
conjectured that it is an upper bound for the length of the
shortest synchronizing word for any $n$-state complete DFA.
The problem can be reduced to automata with strongly connected graph
 \cite{Ce}. The best known upper bound is now equal to $(n^3-n)/6$ \cite{Fr},
\cite{KRS}, \cite{KW}, \cite{Pin}. The conjecture holds true for a lot
of automata, but in general the problem remains open.
This simply looking conjecture is now one of the most longstanding
 open problems in the theory of finite automata.
Moreover, the examples of automata with shortest synchronizing word of
length $(n-1)^2$ are infrequent. After the sequence found by
\v{C}erny and example of \v{C}erny, Piricka and Rosenauerova \cite{CPR}
 of 1971 for $q=2$, the next such example was found by Kari \cite {Ka} only
in 2001 for $n=6$ and $q=2$.  Roman \cite {Ro}
had found an analogical example for $n=5$ and $q=3$ in 2004.
There are no examples of automata for the
time being such that the length of the shortest synchronizing word
is greater than $(n-1)^2$.
\\
 The testing of synchronizing automata is an indispensable part
of investigation in this area \cite{ACV}, \cite{Ep}, \cite{LY}, \cite{Na},
 \cite{RSP}, \cite{Tb}.
The best known to date algorithm of Eppstein \cite{Ep}, \cite{LY}
improves an algorithm of Natarjan \cite{Na} and finds a synchronizing word
for $n$-state DFA in $O(n^3+n^2q)$ time.
\\
We present a new efficient algorithm for finding a synchronizing word.
The actual running time of the algorithm on a lot of examples proved to be
essentially less than in case of $O(n^3q)$ time complexity.
For clear majority of automata, the time complexity is $O(n^2q)$.
 It gives a chance to extend noticeably the class of considered DFA.
This algorithm plays a central role
in the program for search of automata with minimal reset word.
\\
The program studied
all automata with strongly connected transition graph of size $n \le 10$ for $q=2$
and of size $n \le 7$ for $q \le 4$. All known and
some new examples of DFA with shortest synchronizing word of length $(n-1)^2$
from this class of automata were checked.
So all examples of DFA with shortest synchronizing word of length $(n-1)^2$
in this area are known for today. The size of the alphabet of the examples is
two or three. The situation in the neighborhood of the bound $(n-1)^2$ of \v{C}erny
 (minimal reset words of relatively great length) was also studied.
\\
There are no contradictory examples for the \v{C}erny conjecture in this class
of automata.  Moreover, the program does not find new examples
 of DFA with reset word of length $(n-1)^2$ for automata
with $n>4$ as well as for $q>3$. No such examples exist for alphabet of size four
if $n \le 7$.
\\
 And what is more, the examples with minimal length of reset word disappear
 even for values near the \v{C}erny bound $(n-1)^2$ with growth of the size of the
automaton as well as of the size of the alphabet. The gap between $(n-1)^2$
and the nearest of the minimal lengths of reset word appears for $n=6$.
 There are no $6$-state automata with minimal length of synchronizing word
of 24 for $q \le 4$.
\\
 The following table displays this interesting trend
for the length of minimal reset words less than $(n-1)^2$.
\\
\begin{tabular}{|c|c|c|c|c|c|c|}
  \hline
size &   n=5 $q<=4$&n=6 $q<=4$&n=7 $q<=4$&n=8 q=2&n=9 q=2& n=10 q=2 \\
  \hline
$(n-1)^2$    & 16      & 25       & 36       & 49    & 64     & 81 \\
max length   & 15      & 23       & 32       & 44    & 58     & 74  \\
  \hline
\end{tabular}
\\
\\
The program uses also straightforward algorithm for finding synchronizing
word of minimal length. A help algorithm of the program verifies whether
 or not a given DFA is synchronizing.
It is a modification of an algorithm of $O(n^2q)$ time complexity
supposed by Eppstein \cite{Ep}, \cite{LY}. Our version has
$O(n^2q)$ time complexity only in the worst case and we use usually
only its linear part.
\\
 The comparison of the experimental data suggests
that the length of the synchronizing word found by central algorithm of the program
is not far from the length of the minimal synchronizing word. This length
 was not greater than $n^2$ in all billions cases studied for today.
The results of the algorithms altogether correspond
to the \v{C}erny conjecture. All above algorithms are implemented
in our package TESTAS \cite{Tb}.
\\
 \section*{Preliminaries}
Let us consider a deterministic finite automaton with state
transition graph $\Gamma$ and transition semigroup $S$ over
 alphabet $\Sigma$.
The states of the automaton are considered below as vertices of the
transition graph $\Gamma$.
\\
The number of vertices of the graph $\Gamma$ is denoted by $|\Gamma|$.
\\
A maximal strongly connected component of a directed graph will be denoted
 for brevity as $\bf SCC$.
\\
 If there exists a path $v \in \Sigma^+$ from vertex ${\bf p}$
to vertex $\bf q$ in the transition graph of $DFA$ then let us denote
the vertex $\bf q$ as ${\bf p}v$.
\\
Let $\Gamma v$ denote the mapping of the graph [automaton] $\Gamma$ by help of
 $v \in \Sigma^+$, let us call $|\Gamma v|$ {\it rank} and
$|\Gamma| - |\Gamma v|$ {\it defect} of the mapping $v$.
\\
 A word $v \in \Sigma^+ $ is called {\it synchronizing word}
 of an automaton $A$ with transition graph $\Gamma$ if $|\Gamma v|=1$.
 An automaton (and its transition graph) possessing a synchronizing word
is called {\it synchronizing}.
\\
A word $w$ is called {\it 2-reset word}
 of the pair ${\bf p},{\bf q}$ if ${\bf p}w = {\bf q}w$.
\\
Suppose ${\bf p} \succeq \bf q$ if ${\bf p}w ={\bf q}$
for some word $w$.
\\
A state [a vertex] $\bf q$ is called {\it sink} of an
automaton [of a graph] if $\bf p \succeq \bf q$ for all $\bf p$.
 \\
  An automaton [a graph $\Gamma$] is called
{\it complete} if for every state [vertex] $\bf p$
and every $\sigma \in \Sigma$ the state [vertex] ${\bf p}\sigma$ exists.
 \\
The direct product $\Gamma^2$ of two copies of graph $\Gamma$ over
an alphabet $\Sigma$ consists of vertices $({\bf p},{\bf q})$ and
edges (${\bf p},{\bf q}) \to ({\bf p}\sigma,{\bf q}\sigma)$
labelled by $\sigma$. Here ${\bf p},{\bf q} \in \Gamma$, $\sigma
\in \Sigma$.
\section{Some auxiliary properties}
 Two following two simple lemmas belong rather to the folklore.
\begin{lem}  \label {alg} \cite{Ce} \cite{Tt}, \cite{LY}
The directed labelled graph $\Gamma$ is synchronizing if and only
if $\Gamma^2$ has sink state.
 \end{lem}
\begin{lem}  \label {l1} \cite{Tt}
The sets of synchronizing words of the graphs $\Gamma$ and $\Gamma^2$  coincide.
 \end{lem}

 \begin{lem}  \label {l5}
Suppose ${\bf p} \not\in \Gamma s$ for a word $s$ and a state ${\bf p}$
of transition graph $\Gamma$ of DFA.
\\
Then there exist two minimal integer $k$ and $r$ such that
${\bf p}s^k = {\bf p}s^{k+r}$. The pair of states ${\bf p}, {\bf p}s^{r}$
has $2$-reset word $s^{k}$ and for every $i<k$ the pair of states
${\bf p}s^i, {\bf p}s^{r+i}$ has $2$-reset word $s^{k-i}$.
The word $s^{k}$ is a $2$-reset word for at least $k$ different pairs of states.
\\
In the case $r=1$ every pair of states ${\bf p}s^i, {\bf p}s^{k}$
for every $i<k$ has $2$-reset word $s^{k-i}$.
 \end{lem}
Proof. The sequence ${\bf p}s, {\bf p}s^{2}, ...,{\bf p}s^t,... $ is finite
and belongs to $\Gamma s$. Therefore
such $k$ and $r$ exist. Two states ${\bf p}s^i$ and ${\bf p}s^{r+i}$
are mapped by the power $s^{k-i}$ on ${\bf p}s^k = {\bf p}s^{k+r}$
as well as the states ${\bf p}$ and ${\bf p}s^{r}$ are mapped by the
power $s^{k}$ on ${\bf p}s^k$.
All states ${\bf p}s^i$ are distinct for $i \le k$, whence the word $s^{k}$
unites at least $k$ distinct pairs of states.
\\
In the case $r=1$, two states ${\bf p}s^i$ and ${\bf p}s^{k}$
are mapped by the word $s^{k-i}$ on ${\bf p}s^k = {\bf p}s^{k+1}$
as well as the pair of states ${\bf p}$, ${\bf p}s^{k}$ is mapped
by the power $s^{k}$ on ${\bf p}s^k$.
All states ${\bf p}s^i$ are distinct for $i \le k$, whence the word $s^{k}$
unites also in this case at least $k$ distinct pairs of states.
\\
\begin{lem}  \label {l6}
Suppose ${\bf r}\alpha={\bf t}\alpha$ for a letter $\alpha$ and two distinct states
 ${\bf r}$, ${\bf t}$ of transition graph $\Gamma$ of DFA and let the states
${\bf r}$  and ${\bf r}\alpha$ be consecutive states of a cycle $C$ of $\Gamma$.
\\
Then there exists a word $s$ of length of the cycle $C$ such that ${\bf r}s ={\bf r}$
and $|\Gamma s| < |\Gamma|$.
For some state ${\bf p} \in \Gamma \setminus \Gamma s$ there exists a minimal integer
$k$ such that ${\bf p}s^k = {\bf p}s^{k+1}$. The pair of states ${\bf p}, {\bf p}s^{k}$
has $2$-reset word $s^{k}$ and for every $i<k$ the pair of states
${\bf p}s^i, {\bf p}s^{k}$ has $2$-reset word $s^{k-i}$.
The word $s^{k}$ unites at least $k+1$ distinct states.
 \end{lem}
Proof. A word $s$ with first letter $\alpha$ can be obtained
 from consecutive letters on the edges of the cycle $C$.
 Therefore $|s|$ is equal to the length of the cycle
and ${\bf r}s ={\bf r}$.
$|\Gamma s| < |\Gamma|$ follows from ${\bf r}\alpha={\bf t}\alpha$.
\\
From ${\bf r}s ={\bf r} \ne \bf t$ and ${\bf r}\alpha={\bf t}\alpha$ follows
that ${\bf t}s = {\bf r} \ne {\bf t}$ and ${\bf t}s^i \ne {\bf t}$
for any integer $i$. In the case ${\bf t} \in \Gamma \setminus \Gamma s$ suppose
${\bf p} ={\bf t}$, and so the state $\bf p$ is defined. In opposite case for some
state ${\bf t}_1$ holds ${\bf t}_1s ={\bf t}$.
If ${\bf t}_1 \in \Gamma \setminus \Gamma s$ suppose
${\bf p} ={\bf t}_1$, else for some state ${\bf t}_2$ holds
${\bf t}_2s^2 ={\bf t}$. Let us continue this procedure until
${\bf t}_{k-1} \in \Gamma \setminus \Gamma s$ for some $k$
such that ${\bf t}_{k-1}s^{k-1} ={\bf t}$.
Such minimal $k$ exists and all states ${\bf t}$, ${\bf t}_1$, ..., ${\bf t}_j$
 for $j \le k$ are distinct
because ${\bf t}s^i \ne {\bf t}$ for any integer $i$.
The state ${\bf t}$ therefore has a preimage ${\bf p}={\bf t}_{k-1}$ in
$\Gamma \setminus \Gamma s$ by mapping $s^{k-1}$, whence
${\bf p}s^k = {\bf p}s^{k+1}= \bf r$.
\\
So the pair of states ${\bf p}, {\bf p}s^{k}$ has $2$-reset word
$s^{k}$ and for every $i<k$ the pair of states ${\bf p}s^i, {\bf
p}s^{k}$ has $2$-reset word $s^{k-i}$. The states ${\bf p}s^i$ for
$i \le k$ and $\bf p$ are distinct because of the choice of $k$.
The word $s^k$ maps all these states on the state $\bf r$.
\\
\\
Obvious is the following
 \begin{lem}  \label {l7}
Suppose ${\bf q}s =\bf q$ for $m$ states $\bf q$ from $\Gamma$ and
 for some word $s$ such that $s^k =s^{k+1}$.
Then $|\Gamma s^k| = m$.
 \end{lem}
\section{Synchronizing Algorithms}
The following help construction was supposed by Eppstein \cite{Ep}.
Let us keep for any pair of states $\bf r, q$ the first letter $\alpha$ of the
minimal $2$-reset word $w$ of the pair of states together
with the length of the word $w$. The corresponding letter
of the pair of states ${\bf r}\alpha,{\bf q}\alpha$ is the second letter of $w$.
The $2$-reset word $w$ of minimal length can be restored on this way.
 The time and space complexity of this preprocessing is $O(|\Gamma^2|)$ \cite{Ep}
and it will be used in majority of considered algorithms.
\\
A help algorithm with $O(|\Gamma|^2q)$ time complexity in the
 worst case based on Lemmas \ref{alg} and \ref{l1} verifies whether or
not a given DFA is synchronizing \cite{Ep}, \cite{Tb}.
The main part of the algorithm follows \cite{Ep} (see also \cite{LY}).
Our modification of the algorithm finds first all SCC of the graph
(a linear algorithm) and then checks the minimal SCC of the graph
(if exists). The program for search
of automata with relatively great minimal reset word uses this
algorithm on the preliminary (and quite often linear) stage.
\\
 An efficient semigroup algorithm, essential improvement of
the algorithm from \cite{Ep}, based on the properties of syntactic
 semigroup and inspired by Lemmas \ref{l5} - \ref{l7} is used on
the next stage and plays a central role in the program.
 \subsection{A semigroup algorithm for synchronizing word}
We consider the square $\Gamma^2$ and the reverse graph $I$ of $\Gamma$.
The graph $I$ is not deterministic for synchronizing graph $\Gamma$.
\\
Suppose that the graph $\Gamma$ is synchronizing, all sink states
are found on the stage of checking of the synchronizability,
 the graph $\Gamma^2$ and the reverse graph $I$ were build.
\\
Let us find by help of the reverse graph $I$ for any pair of states
$\bf r, q$ from $\Gamma^2$ the first letter of the minimal $2$-reset word
$w$ of the pair and the length of $w$ \cite{Ep}.
So for any pair of states ($\bf r, q$) can be restored a
$2$-reset word $w$ of minimal length.
\\
The set of states ($\bf r, q$) can be ordered according to the length
of the word $w$. The ordering can be made linear in the size of the set.
One can find first the number of all pairs ($\bf r, q$)
 with given length of minimal $2$-reset word for any length,
then adjust an interval for to place the pairs and then allocate
the pairs of states in the interval according to the value of the
length.
\\
We use also an another idea for to reorder the pairs
 of states. The number of preimages
of the state ${\bf r}w={\bf q}w$ by mapping $w^k$ for any integer $k$
 can be used for the ordering together with the length $|w|$.
Let us call this order {\it the second}.
The number of preimages can be found
in linear time for given pair of states ($\bf r, q$) using the reverse graph $I$.
The corresponding words may form a set of generators of a subsemigroup of the
semigroup $A$ of all reset words and we will use only linear number of pairs studied
for this aim.
\\
The important part of the preprocessing supposed by Eppstein was
the computing of the mapping $\Gamma w$ of the graph $\Gamma$
induced by the minimal $2$-reset word $w$ of the pair of states $\bf r, q$.
This stage begins from the shortest words $w$ and therefore is linear
for any considered pair of states $\bf r, q$. Nevertheless, the time
complexity of the stage is $O(\Gamma^3)$.
 For to avoid the extremes of this step, our algorithm stops on linear number of
pairs.
  The obtained set $G$ of $2$-reset words is considered as a set of generators of
some subsemigroup from $A$ and will be marked together with corresponding pairs
of states. The time complexity of this step is therefore $O(\Gamma^2)$.
Let us reorder $G$ in the {\it second} order and use
the mapping of the graph induced  by powers of generators.
\\
Let $\Gamma_i$ be consecutive images of the
 graph $\Gamma=\Gamma_0$ such that for $w_i \in A$ holds
$\Gamma_iw_{i+1}=\Gamma_{i+1}$ and $|\Gamma_i| >|\Gamma_{i+1}|$.
Let $A_i$ be a semigroup generated by the set $w_1$, ... $w_i$.
Let us check pairs of states corresponding to the words from $G$.
If the pair belongs to $\Gamma_i$ then the corresponding minimal reset
word $w_{i+1}$ may be used for to find the image $\Gamma_{i+1}$.
\\
In the case no minimal $2$-reset word of a pair from $\Gamma_i$
was marked, let us consider the products of marked words.
If some product unites a pairs of states of $\Gamma_i$, then let us use the
mapping, mark the product of words and the pair of states.
Let us notice that on this step are considered not all marked pairs.
The number of considered products must be linear in the size of $\Gamma$.
The product of two mappings can be found in linear time.
Therefore the time complexity of this stage is $O(|\Gamma| k)$ for the defect $k$
of the mapping of $\Gamma_i$.
\\
If two considered stages still do not find a reset word,
 then the new generator must be added
to considered subsemigroup $A_i$. Let us take a pair of
states $\bf r, q$ from $\Gamma_i$ with reset word $w_i$.
Suppose $w_i= u_iv_i$ such that the word $v_i$ was
marked. Then the mapping $w_i$ can be found in
$|\Gamma| |u_i|$ time. Let us notice that only on this step
the time complexity may by greater than quadratic.
\begin{lem}  $\label {2}$
Let $\Gamma_i$ be consecutive images of the
 graph $\Gamma=\Gamma_0$ such that for $v_i$ from semigroup $A$
$\Gamma_iv_{i+1}=\Gamma_{i+1}$, $|\Gamma_i| >|\Gamma_{i+1}|$ and $|\Gamma_s| =1$
 for some integer $s$.
Let $A_i$ be a semigroup generated by the set $w_1$, ... $w_i$ such that $w_i=u_iv_i$
is a reset word for some pair of states from $\Gamma_{i-1}$ and
  $v_i$ is a marked element of the subsemigroup $A_{i-1}$.
\\
Then the  considered algorithm has $max(O(|\Gamma|^2q), O(|\Gamma||u_1...u_s|)$
time complexity.
 \end{lem}
\begin{proof}
The time complexity of the step of the building of $\Gamma^2$ is
$O(|\Gamma|^2q)$. So $O(|\Gamma|^2q)$ is a lower bound for the complexity
of the considered algorithm.
\\
Let the set $w_1$, ... $w_i$ generate $A_i$. The creation of the mapping
$w_i$ needs $|\Gamma||u_i|+1$ steps because for the marked element $v_i$
the mapping is known.
\\
The element will be marked and used only if it is either a generator
from $A_i$ or a product of two marked elements.
 With a marked semigroup element will be associated
the mapping of $\Gamma$ defined by the element.
The finding of the mapping of the product of two elements with known images
is linear in the size of the graph.
\end{proof}
 We repeat the process with the obtained image  $\Gamma_i$.
The defect of the mapping is growing on every step.
After not over than $|\Gamma|-1$ steps
$\Gamma$ will be synchronized.
\\
The process of recording of the synchronizing word is linear in
the length of the word. The length of the synchronizing word
found by the algorithm in billions of practical experiments was
less than $|\Gamma|^2$ in all considered cases.
The stage of adding of new generators was used only in a small
number of cases, only some percents of considered synchronizing
automata. The minimal number of generators of the semigroup $A$
is usually small. For instance, for all \v{C}erny graphs there are only
two generators.
Therefore the time complexity of the algorithm
is $O(|\Gamma|^2q)$ in overwhelming majority of cases and the
algorithm can be considered as almost quadratic.
\subsection{Modification of Eppstein algorithm}
Some version of the program uses also a modification of Eppstein algorithm
 \cite{Ep}, \cite{LY} for finding synchronizing word of
$O(|\Gamma|^3 + (|\Gamma|^2q)$ time complexity. The favorable idea
of Eppstein was to keep with any pair of states $\bf r, q$
the first letter of the minimal reset word $w$, its length and the image of the
set of states by help of the mapping induced by the word $w$.
The building of the images has $O(|\Gamma|^3)$ time complexity
and is a most wasteful part of the algorithm.
\\
Our modification of the Eppstein algorithm (called below
a cycle algorithm) instead of a word $w$ considers a power of this
 word until stabilization of the rank of the image.
It proved to be fruitful in many cases including such extraordinary case
as graphs of \v{C}erny \cite {Ce}. The length of the reset word obtained
 by the algorithm in this case reaches its minimum. We omit sometimes
this stage of the program despite the growing number of the
graphs studied on the next stage. Nevertheless, the
observation period of the whole of the program is essentially smaller
in spite of the fact that the next stage is non-polynomial.
\\
\begin{thm} \cite{Fr}, \cite{KRS} \label{kf}
Let $C$ be set of size $k$ and let us consider a sequence
 of its subsets $C_i$ of size $m$ such that any $C_i$ includes a
two-element subset of $C$ not included in every $C_j$ for $j<i$.
Then the length of the sequence is less than
$(k-m+2)*(k-m+1)/2$.
\end{thm}
\begin{cor} \label{kf1}
Let $\Gamma$ be transition graph of an automaton with $|\Gamma|$ states
and let us consider a sequence
 of subsets $C_i$ of states of the automaton of size $m$ or less
  such that any $C_i$ includes a two-element subset of states
  of $\Gamma$ not included in every $C_j$
for $j<i$. Suppose the length of the sequence is
$(|\Gamma|-m+2)*(|\Gamma|-m+1)/2$. Then at least one $C_i$ contains less than
$m$ states.
 Any sequence of length $(|\Gamma|^3-|\Gamma|)/6$ of considered kind for distinct
 $m$ contains a set of size one.
\end{cor}
The value $(|\Gamma|^3-|\Gamma|)/6$ is well known and was mentioned time and again
 \cite{Fr}, \cite{KRS}, \cite{KW}, \cite{Pin}.
The combinatorial theorem \ref{kf}
can be used for estimation of the length of the reset word obtained by Eppstein,
cycle and semigroup algorithms.
 The theorem considers distinct mappings of the graph of the
automaton induced by the letters of the alphabet of the labels
such that any new mapping has at least one pair of states that does
not belong to any previous mapping of the same rank. For given rank $k$ of
mapping in considered algorithms there are at most
$(|\Gamma|+k)(|\Gamma| + k - 1)/2$ or less than
$|\Gamma|$ distinct mappings. The pair of states
with a most short reset word creates a sequence of such mappings and therefore
the theorem \ref{kf} can be used here.
 Corollary \ref{kf1} implies
\begin{prop}
The length of the reset word obtained by Eppstein, cycle and semigroup algorithms
is less than $(|\Gamma|^3-|\Gamma|)/6$.
\end{prop}
So the time complexity of the algorithm in the most worst case is $O(|\Gamma|^3q)$.
Really this most worst case is very rare, for all automata studied for today by
these algorithms, it was less than $|\Gamma|^2$.
\subsection{An algorithm for finding synchronizing word of minimal
length}
On the last stage, the program uses a straightforward algorithm for finding
synchronizing word of minimal length. The last one is not polynomial in the
most worst case (the finding of the synchronizing
word of minimal length is NP-hard \cite{Ep}, \cite{LY}, \cite{Sa}).
The program for search of minimal reset word uses this
algorithm relatively rare.
\\
 The algorithm is a revision of an algorithm for finding
the syntactic semigroup $S$ of size $s$ with $q$ generators on the base of
transition graph \cite{TW}. We find mappings of the
graph of the automaton induced by the letters of the alphabet of
the labels. Mappings with the same set
 of states are identified. It essentially simplified  the process
in comparison with the algorithm from \cite{TW}.
 Distinct mappings are saved. For this aim, any two mappings must
to be compared, so we have $O(s(s-1)/2)$ steps. Let us notice that
the size of the syntactic semigroup is in general not polynomial
in the size of the transition graph.
\\
The mappings correspond to semigroup
 elements. With any mapping let us connect a previous mapping and the letter
that creates the mapping. On this way, the path on the graph of the automaton
  can be constructed.
 \begin{prop}  \label {propl}
The algorithm finds a list of all words (elements of syntactic
semigroup) of length $k$ where $k$ is growing. The first
synchronizing word of the list has minimal length.
\end{prop}
The time complexity of the considered procedure is
$O(|\Gamma|qs^2)$ with $O(|\Gamma|s)$ space complexity.
\subsection{Checking synchronizability}
The algorithm is based on the Lemma \ref{alg} and presents a
modification of an algorithm from \cite{Ep}.
\\
First let us check SCC using the first-depth search and find the
SCC $\Gamma_s$ of sink states from $\Gamma$. If there are no sink
state then the graph has no synchronizing word and the algorithm
stops.  Exactly one sink state implies synchronizability and the
algorithm also stops. The time and space complexity of these step
are linear. Now we can consider the graph $\Gamma_s$ with at least
two sink states.
\\
The next step is the consideration of $\Gamma_s^2$.
We unite any pair of states ($\bf p, q$) and ($\bf q, p$), all states
($\bf r, r$) are united in one state ($0, 0$).
Then let us mark sink state ($0, 0$) and all ancestors of ($0, 0$)
 using the first-depth search on the reverse of the obtained graph $G$.
The graph $\Gamma$ is synchronizing if any node of $G$ will be marked.
 The time and space complexity of the algorithm in the most worst case is
$O(|\Gamma|^2q)$.
\section{Experimental data}
The considered synchronization algorithms were used in a program
for search of automata with minimal reset word of relatively great
length. The program has investigated all complete DFA for $n \le
10$, $q=2$ and for $n \le 7$, $q \le 4$.
\\
An automaton with $k$ states outside sink $SCC$ $A$ of the
transition graph can be mapped on $A$ by word of length not
greater than $k(k-1)/2$. Therefore only automata with strongly
connected transition graphs need investigation. The graphs with
synchronizing proper subgraph obtained by moving off letters from
the alphabet are omitted too. The program reduced also the number
of studied isomorphic copies of automata. The case of $n=2$ is not
considered because any synchronizing automaton with two states has
reset word of length $(n-1)^2=1$.
\\
The known $n$-state automata with minimal
 reset word of length $(n-1)^2$ are presented by sequence of
 \v{C}erny \cite{Ce} (here n=28):
 \\
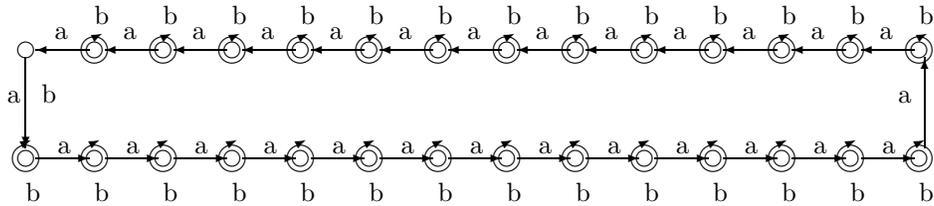
\begin{picture}(300,73)
\multiput(0,54)(26,0){14}{\circle{6}}
\multiput(26,54)(26,0){13}{\circle{10}}
\multiput(24,54)(26,0){13}{\vector(-1,0){20}}
\multiput(11,58)(26,0){13}{a}
 \multiput(26,64)(26,0){13}{b}
 \multiput(25,58)(26,0){13}{\vector(2,1){4}}
\put(340,17){\vector(0,1){34}}
 \put(0,51){\vector(0,-1){34}}
\put(-7,34){a} \put(6,34){b} \put(330,34){a}
\multiput(0,13)(26,0){14}{\circle{6}}
\multiput(0,13)(26,0){14}{\circle{10}}
\multiput(4,13)(26,0){13}{\vector(1,0){20}}
\multiput(12,15)(26,0){13}{a}
 \multiput(0,-3)(26,0){14}{b}
 \multiput(-2,18)(26,0){14}{\vector(2,1){4}}
 \end{picture}
\\
\\
 by automata supposed by \v{C}erny, Piricka
and Rosenauerova \cite{CPR} ($CPR$), by Kari \cite{Ka} and Roman \cite{Ro}.
\\
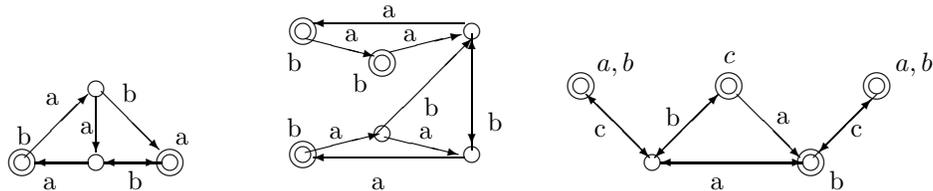
\begin{picture}(100,60)
\multiput(2,10)(28,0){3}{\circle{6}}
\multiput(2,10)(56,0){2}{\circle{10}}
\put(0,17){b}
\put(60,17){a}
 \put(30,38){\circle{6}}
\put(27,10){\vector(-1,0){20}}
\put(33,10){\vector(1,0){20}}
\put(55,10){\vector(-1,0){22}}
\put(3,12){\vector(1,1){24}}
\put(32,37){\vector(1,-1){22}}
\put(30,35){\vector(0,-1){21}}
 \put(40,33){b}
\put(10,0){a}
  \put(42,0){b}
\put(11,32){a}
\put(24,21){a}
 \end{picture}
\begin{picture}(130,78)
\multiput(6,60)(64,0){2}{\circle{6}}
\multiput(6,13)(64,0){2}{\circle{6}}
 \multiput(22,56)(22,0){2}{a}
\multiput(16,19)(34,0){2}{a}
 \put(36,21){\circle{6}}
\put(36,48){\circle{6}}
 \put(7,14){\vector(4,1){28}}
\put(7,57){\vector(4,-1){26}}

\put(39,52){\vector(4,1){27}}
 \put(37,20){\vector(4,-1){28}}
\put(67,63){\vector(-1,0){57}}
 \put(36,65){a}
\put(67,12){\vector(-1,0){57}}
 \put(32,0){a}

\put(70,15){\vector(0,1){42}}
 \put(70,59){\vector(0,-1){42}}
\put(34,21){\vector(1,1){36}}
 \put(52,28){b}

  \put(76,22){b}

\put(25,37){b}
\put(36,48){\circle{10}}
\put(0,20){b}
\put(0,45){b}

\put(6,60){\circle{10}}
 \put(6,13){\circle{10}}
 \end{picture}
\begin{picture}(150,80)
\multiput(-21,39)(56,0){3}{\circle{6}}
\multiput(-21,39)(56,0){3}{\circle{10}}

 \multiput(6,10)(60,0){2}{\circle{6}}

\put(66,10){\circle{10}}
 \put(5,12){\vector(-1,1){24}}
  \put(-19,36){\vector(1,-1){24}}

\put(67,12){\vector(1,1){22}}
  \put(91,36){\vector(-1,-1){22}}

  \multiput(-16,20)(97,0){2}{c}

\put(7,12){\vector(1,1){24}}
\put(31,36){\vector(-1,-1){24}}

\put(38,36){\vector(1,-1){24}}

   \put(9,10){\vector(1,0){54}}
   \put(63,10){\vector(-1,0){54}}

 \put(28,0){a}
\put(53,25){a}
 \put(11,24){b}

\put(-15,45){$a,b$}
\put(33,48){$c$}
\put(98,45){$a,b$}
\put(73,0){b}
 \end{picture}
\\
Our program has found five new following examples on the border $(n-1)^2$.
The loops of the complete graphs are omitted here for simplicity.
\\
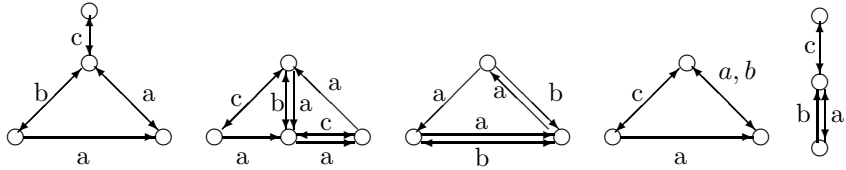
\begin{picture}(72,65)
\multiput(6,10)(56,0){2}{\circle{6}}
 \put(34,38){\circle{6}}
\put(34,58){\circle{6}}
 \put(9,10){\vector(1,0){50}}
\put(7,12){\vector(1,1){24}}
\put(31,36){\vector(-1,-1){24}}
\put(36,37){\vector(1,-1){24}}
\put(60,13){\vector(-1,1){24}}
\put(34,41){\vector(0,1){15}}
\put(34,56){\vector(0,-1){15}}
 \put(13,24){b}
\put(29,0){a}
  \put(54,24){a}
\put(27,45){c}

 \end{picture}
\begin{picture}(72,60)
\multiput(6,10)(28,0){3}{\circle{6}}
 \put(34,38){\circle{6}}
 \put(9,10){\vector(1,0){22}}
\put(37,8){\vector(1,0){22}}
\put(37,11){\vector(1,0){22}}
\put(59,11){\vector(-1,0){22}}

\put(60,13){\vector(-1,1){24}}
\put(32,36){\vector(-1,-1){23}}
\put(9,13){\vector(1,1){23}}
\put(36,35){\vector(0,-1){22}}

\put(33,35){\vector(0,-1){22}}
\put(33,14){\vector(0,1){21}}

 \put(49,28){a}
\put(14,0){a}
  \put(46,0){a}
\put(46,13){c}
\put(12,22){c}

\put(38,20){a}
\put(27,20){b}

 \end{picture}
\begin{picture}(72,60)
\multiput(6,10)(56,0){2}{\circle{6}}
 \put(34,38){\circle{6}}
 \put(9,11){\vector(1,0){50}}
\put(29,13){a}

\put(59,8){\vector(-1,0){50}}
\put(9,8){\vector(1,0){50}}
\put(29,-1){b}

\put(31,36){\vector(-1,-1){24}}
\put(37,37){\vector(1,-1){24}}
\put(57,24){b}

\put(36,25){a}
\put(57,13){\vector(-1,1){22}}
 \put(13,24){a}
 \end{picture}
\begin{picture}(75,60)
\multiput(6,10)(56,0){2}{\circle{6}}
 \put(34,38){\circle{6}}
 \put(9,10){\vector(1,0){50}}
\put(7,12){\vector(1,1){24}}
\put(31,36){\vector(-1,-1){24}}
\put(36,37){\vector(1,-1){24}}
\put(60,13){\vector(-1,1){24}}
 \put(13,24){c}
\put(29,0){a}
  \put(46,32){$a,b$}

 \end{picture}
\begin{picture}(12,60)
\multiput(6,6)(0,25){3}{\circle{6}}
 \put(5,8){\vector(0,1){20}}
\put(-3,16){b}

\put(8,28){\vector(0,-1){19}}
\put(8,8){\vector(0,1){20}}
\put(10,16){a}

\put(6,53){\vector(0,-1){19}}
\put(0,42){c}

\put(6,34){\vector(0,1){20}}
 \end{picture}
\\
The corresponding reset words of minimal length are: $\it
abcacabca$,  $\it acbaaacba$, $\it baab$, $\it acba$, $\it bacb$.
All considered algorithms have found the same reset word for every
example. The size of the syntactic semigroup found by the package
TESTAS is 148, 180, 24, 27 and 27 correspondingly.
\\
No doubts that some automata from this list, especially for $n=3$, were
sometimes studied by specialists, but we have not found any mention
of.
\\
There are no contradictory examples for the \v{C}erny conjecture in considered class
of automata. Moreover, the program does not find new examples of automata with
reset word of length $(n-1)^2$ for $n>4$ and $q>3$.
\\
 And what is more, the examples with minimal length of reset word disappear
 even for values near the \v{C}erny bound $(n-1)^2$ with growth of the size of the
automaton. The gap appears for $n=6$. There are no $6$-state
automata with minimal length of synchronizing word equal to 24 for
$q \le 4$.
\\
 The following table displays this noteworthy trend
for the maximum of lengths of minimal
 reset words. The mentioned above examples on the \v{C}erny border are not
taken in account in the third line of the table.
 \\
\\
\begin{tabular}{|c|c|c|c|c|c|c|}
  \hline
size &   n=5 $q<=4$&n=6 $q<=4$&n=7 $q<=4$&n=8 q=2&n=9 q=2& n=10 q=2 \\
  \hline
$(n-1)^2$    & 16      & 25       & 36       & 49    & 64     & 81 \\
max length   & 15      & 23       & 32       & 44    & 58     & 74  \\
  \hline
\end{tabular}
\\
\\
The gap between $(n-1)^2$ and the length of the minimal reset word
 grows with $n$.
This growing gap supports the following funny
 \begin{con} The set of $n$-state DFA $(n>2)$ with minimal reset word
of length $(n-1)^2$ contains only the sequence
of \v{C}erny and the eight automata mentioned above, three of size 3,
 three of size 4, one of size 5 and one of size 6.
\end{con}
Let us consider the synchronization algorithms from the package
TESTAS on some above-mentioned objects and on
  a modification \cite{Tr} of a graph KMM supposed by Kim, McNaughton,
 McCloskey \cite {K91}.
\\
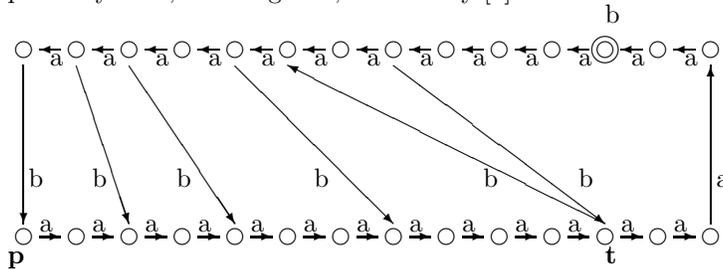
\begin{picture}(300,100)
\multiput(6,84)(20,0){14}{\circle{6}}
\multiput(20,84)(20,0){13}{\vector(-1,0){8}}
\multiput(16,78)(20,0){13}{a}

\put(266,18){\vector(0,1){60}}
\put(6,78){\vector(0,-1){60}}
\put(8,32){b}
\put(268,32){a}

\put(26,78){\vector(1,-3){20}}
\put(32,32){b}
\put(46,78){\vector(2,-3){40}}
\put(64,32){b}
\put(86,78){\vector(1,-1){60}}
\put(116,32){b}
\put(146,78){\vector(4,-3){80}}
\put(180,32){b}
\put(226,18){\vector(-2,1){120}}
\put(216,32){b}

\put(226,84){\circle{10}} \put(226,94){b}
\put(0,3){\bf p}
\put(226,3){\bf t}

\multiput(6,13)(20,0){14}{\circle{6}}
\multiput(12,13)(20,0){13}{\vector(1,0){8}}
\multiput(12,15)(20,0){13}{a}
 \end{picture}
 \\
  Complete closure KMML of this graph is obtained from KMM by adding loops
in all necessary cases. The $n$-state automata supposed by
\v{C}erny will be denoted by C$<n>$.
\\
The following table presents the name of the automaton,
    the number of its states,
 the size of the syntactic semigroup, the length of synchronizing word found
 by the Eppstein algorithm \cite{Ep},  by the cycle and the semigroup algorithm,
 by the minimal synchronizing word algorithm with the
 corresponding number of mappings of the set of states.
\\
 \\
\begin{tabular}{|c|c|c|c|c|c|c|c|c|c|c|c|c|c|}
  \hline
    name & CPR & Roman &  Kari  & C6 & C9 & C17 & KMM & KMML & C28 & C151\\
  \hline
graph size  & 4   &  5    &  6    & 6    & 9     & 17  & 28   & 28 & 28 & 151 \\
semigroup size & 145 & 1397  & 17265 & 2742 & 218718 & huge & 22126 & $>10^6$ & huge & huge \\
 Eppstein alg & 9 & 17     & 26    & 27  & 78     & 375 & 4 & 51 & 1202 & 57190\\
cycle algorithm & 9 & 18     & 27    & 25   & 64     & 256 & 4 & 57 & 729 & 22500 \\
semigroup alg & 9 & 17     & 27    & 25  & 64     & 256 & 4 & 27 & 729 & 22500\\
minimal length & 9 & 16    & 25    & 25   & 64     & 256 & 4 & 27 & 729 & 22500 \\
mappings & 9  & 22 & 46    & 56    & 501 & 131053 & 12  &41035 & vast & vast \\
  \hline
\end{tabular}
\\
\\
One can compare the results of the algorithms.
Equality of the length of minimal synchronizing word and of synchronizing
word found by the semigroup algorithm and by Eppstein or cycle
algorithm holds in some cases. In particular, it's true even for
such extreme objects as \v{C}erny automata. Moreover, we obtain
not infrequently the same synchronizing words. The transition
semigroup of the \v{C}erny automaton has a nilpotent element of
order $n-1$, and the minimal synchronizing word of the automaton
is a subword of a power of this element.
\\
 As for the size of the syntactic semigroup from the
table, the most discouraging example gives us the Kari automaton.
The size of the syntactic semigroup of the \v{C}erny automaton is
very great too, it is about $O(2^{2n})$. Maximal size $n^n$ of the
syntactic semigroup is reached for the examples of $n=3$, $q=3$.
It is the semigroup of all transformations of 3-element set.

 \end{document}